\documentclass[12pt,letterpaper]{article}
\usepackage{amsmath,amssymb,array,calc,rotating,epsfig,psfrag,amscd, cite}

\numberwithin{equation}{section}

\makeatletter
\renewcommand\section{\@startsection {section}{1}{\z@}%
                                   {-3.5ex \@plus -1ex \@minus -.2ex}%nn
                                   {2.3ex \@plus.2ex}%
                                   {\normalfont\large\bfseries}}
\renewcommand\subsection{\@startsection{subsection}{2}{\z@}%
                                     {-3.25ex\@plus -1ex \@minus -.2ex}%
                                     {1.5ex \@plus .2ex}%
                                     {\normalfont\bfseries}}
\makeatother

\let\non\nonumber

\let\S=\Sigma

\newcommand{\bea}{\begin{eqnarray}}
\newcommand{\eea}{\end{eqnarray}}
\newcommand{\be}{\begin{equation}}
\newcommand{\ee}{\end{equation}}

\newcommand{\p}{\partial}

\newcommand{\C}[1]{$(\ref{#1})$}

\def\IZ{\relax\ifmmode\mathchoice
{\hbox{\cmss Z\kern-.4em Z}}{\hbox{\cmss Z\kern-.4em Z}}
{\lower.9pt\hbox{\cmsss Z\kern-.4em Z}} {\lower1.2pt\hbox{\cmsss
Z\kern-.4em Z}}\else{\cmss Z\kern-.4em Z}\fi}
\def\IR{\relax{\rm I\kern-.18em R}}

\def\one{{\hbox{ 1\kern-.8mm l}}}

\newlength{\bredde}
\def\slash#1{\settowidth{\bredde}{$#1$}\ifmmode\,\raisebox{.15ex}{/}
\hspace*{-\bredde} #1\else$\,\raisebox{.15ex}{/}\hspace*{-\bredde}
#1$\fi}

\newsavebox{\zzzbar}
\sbox{\zzzbar}
  {\setlength{\unitlength}{0.9em}
  \begin{picture}(0.6,0.7)
  \thinlines
  \put(0,0){\line(1,0){0.6}}
  \put(0,0.75){\line(1,0){0.575}}
  \multiput(0,0)(0.0125,0.025){30}{\rule{0.3pt}{0.3pt}}
  \multiput(0.2,0)(0.0125,0.025){30}{\rule{0.3pt}{0.3pt}}
  \put(0,0.75){\line(0,-1){0.15}}
  \put(0.015,0.75){\line(0,-1){0.1}}
  \put(0.03,0.75){\line(0,-1){0.075}}
  \put(0.045,0.75){\line(0,-1){0.05}}
  \put(0.05,0.75){\line(0,-1){0.025}}
  \put(0.6,0){\line(0,1){0.15}}
  \put(0.585,0){\line(0,1){0.1}}
  \put(0.57,0){\line(0,1){0.075}}
  \put(0.555,0){\line(0,1){0.05}}
  \put(0.55,0){\line(0,1){0.025}}
  \end{picture}}

\newcommand{\ena}{\end{eqnarray}}
\newcommand{\beqa}{\begin{eqnarray}}
\newcommand{\eeqa}{\end{eqnarray}}

\def\S{\Sigma}

\begin{document}
\begin{titlepage}

\begin{center}

%\hfill \today
%\hfill         \phantom{xxx}         

%\hfill HRI

\vskip 2 cm
{\Large \bf Poisson equations for elliptic modular graph functions}\\
\vskip 1.25 cm { Anirban Basu\footnote{email address:
    anirbanbasu@hri.res.in} } \\
{\vskip 0.5cm  Harish--Chandra Research Institute, HBNI, Chhatnag Road, Jhusi,\\
Prayagraj 211019, India}

\end{center}

\vskip 2 cm

\begin{abstract}
\baselineskip=18pt

We obtain Poisson equations satisfied by elliptic modular graph functions with four links. 
Analysis of these equations leads to a non--trivial algebraic relation between the various graphs.           

\end{abstract}

\end{titlepage}

%\pagestyle{plain}
%\baselineskip=18pt
% Try a wider skip
%\baselineskip=19pt
%%%%%%%%%%%%%%%%%%%%%%%%%%%%%%%%%%%%%%%%%%%%%%%%%%%%%%%%%%%%%%%%%%%%%%%%%%%%%%

\section{Introduction}

Modular graph functions, or more generally modular graph forms, play a key role in the analysis of perturbative contributions to higher derivative interactions in the low energy effective action of superstring theory, and have been studied upto genus two in various contexts. The links in these graphs at genus $g$ are given by scalar Green functions or their derivatives on the genus $g$ Riemann surface, while the vertices are integrated over the worldsheet~\cite{DHoker:2015gmr,DHoker:2015wxz} with a non--trivial measure involving the worldsheet moduli. These graphs are $Sp(2g,\mathbb{Z})$ covariant and they contribute to the $Sp(2g,\mathbb{Z})$ invariant integrand, whose integral over the moduli space of the genus $g$ Riemann surface yields coefficients of terms in the effective action that are analytic in the external momenta.       

A non--trivial generalization of these graphs arises when two of the vertices are not integrated over the worldsheet\footnote{All the vertices that arise in the modular graphs mentioned above are integrated. However, using translational invariance at genus one, one of the vertices can be fixed and hence unintegrated.}. At genus one, they have been discussed in~\cite{DHoker:2015wxz} and referred to as single--valued elliptic multiple polylogarithms. Apart from being interesting in their own right, they arise in the asymptotic expansion of genus two modular graph functions around the non--separating node~\cite{DHoker:2017pvk,DHoker:2018mys}. They have been referred to as generalized modular graph functions in~\cite{DHoker:2018mys} and as elliptic modular graph functions in~\cite{DHoker:2020tcq}, which is the terminology we shall use.

A non--trivial algebraic identity between genus two modular graphs was obtained in\cite{DHoker:2020tcq}, which can be analyzed in an asymptotic expansion around the degeneration nodes. It was found that on expanding around the non--separating node, the genus two identity yields a non--trivial Poisson equation involving various graphs at genus one. Each term in this equation involves graphs with a total of four links. Furthermore, some of the graphs have two vertices that are not integrated over and hence are elliptic, while the others simply are modular graphs. For the elliptic modular graphs, let the locations of the two unintegrated vertices be $p_b$ and $p_a$. Using translational invariance on the torus, we can shift them to be at $v= p_b -p_a$ and at 0, which is the convention we shall use. From the point of view of the degenerating genus two Riemann surface, $p_b$ and $p_a$ are the two punctures on the worldsheet at the non--separating node which are connected by a thin long tube, roughly the inverse length of which is the expansion parameter. While this eigenvalue equation has been deduced as a consequence of the genus two identity, it is interesting to obtain it directly from an analysis at genus one.                         

In this paper, we obtain this Poisson equation directly at genus one, without worrying about its genus two origin. To start with, we briefly review relevant definitions and properties of various genus one invariants. We next obtain the Poisson equation by analyzing the variation of the Green function in the various graphs under deformations of the complex structure of the torus. We also obtain the Poisson equation satisfied by a different elliptic modular graph with four links. Analyzing these equations, we obtain a non--trivial algebraic equation relating various elliptic modular graphs as well as modular graphs. Some of the terms in this equation are quadratic in the graphs.        

When the two unintegrated vertices at $v$ and 0 are identified, the elliptic modular graphs reduce to modular graphs, and hence our results generalize those for modular graphs. For such graphs, several Poisson equations as well as algebraic relations between them have been obtained that are of relevance to the low momentum expansion of the effective action of type II string theory~\cite{DHoker:2015gmr,DHoker:2015sve,DHoker:2015wxz,Basu:2015ayg,DHoker:2016mwo,Basu:2016xrt,Basu:2016kli,Basu:2016mmk,DHoker:2016quv,Kleinschmidt:2017ege,DHoker:2019blr,Basu:2019idd,Gerken:2019cxz,Gerken:2020yii,Gerken:2020aju}. Our analysis follows the method of deriving the Poisson equations in~\cite{Basu:2015ayg,Basu:2016xrt,Basu:2016kli,Kleinschmidt:2017ege,Basu:2019idd} by analyzing the variations of the Green functions that arise in the graphs under deformations of the complex structure. We expect our analysis of the elliptic modular graphs along these lines to be generalizable to various cases.         

\section{Genus one modular invariant graphs and their variations under complex structure deformations} 

The links of the various graphs are given by the scalar Green function $G(z,w) = G(z-w)$ on the toroidal worldsheet. The Green function is given by~\cite{Lerche:1987qk,Green:1999pv}
\be \label{Green} G(z) = \frac{1}{\pi} \sum_{(m,n) \neq (0,0)} \frac{\tau_2}{\vert m\tau+n\vert^2}e^{\pi[\bar{z}(m\tau+n)- z(m\bar\tau+n)]/\tau_2}.\ee
It is modular invariant and doubly periodic on the torus. Since it is single valued, we can integrate by parts and neglect total derivatives which proves to be very useful in our calculations. 

Denoting by $z$ the coordinate on the torus, we have that
\be -\frac{1}{2} \leq {\rm Re} z \leq \frac{1}{2} , \quad 0 \leq {\rm Im} z \leq \tau_2.\ee
The measure is given by $d^2 z= d{\rm Re}z d{\rm Im} z$, while the Dirac delta function is normalized to satisfy $\int_{\S}d^2 z \delta^2 (z)=1$. In the integrals over the worldsheet, we denote the worldsheet by $\S$.

\begin{figure}[ht]
\begin{center}
\[
\mbox{\begin{picture}(240,90)(0,0)
\includegraphics[scale=.65]{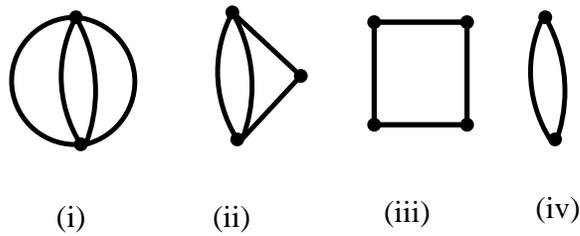}
\end{picture}}
\]
\caption{The modular graphs (i) $D_4$, (ii) $C_{1,1,2}$, (iii) $E_4$, (iv) $E_2$}
\end{center}
\end{figure}

From the definition \C{Green}, it follows that
\be \label{Vanish}\int_{\S} d^2 z G(z,w)=0\ee
which is very useful for our purposes. Thus there can be no graphs where a single link representing a Green function ends on an integrated vertex. 

First let us consider the modular graphs where the ones that are relevant to us are given in figure 1. The first two graphs are given by  
\be \label{defD}D_4 = \int_{\S} \frac{d^2 z}{\tau_2} G(z)^4, \quad C_{1,1,2} = \int_{\S^2} \frac{d^2 z}{\tau_2}\frac{d^2 w}{\tau_2} G(z,w)^2G(z)G(w) ,\ee
while the others are special cases of the non--holomorphic Eisenstein series $E_k$ defined by
\bea E_k &=& \int_{\S^{k-1}} \prod_{i=1}^{k-1} \frac{d^2 z_i}{\tau_2} G(z_1)G(z_1-z_2)\ldots G(z_{k-2}- z_{k-1})G(z_{k-1}) \non \\ &=&\frac{1}{\pi^k} \sum_{(m,n) \neq (0,0)} \frac{\tau_2^k}{\vert m\tau+n\vert^{2k}}\eea
for $k \geq 2$.

The $SL(2,\mathbb{Z})$ invariant Laplacian is given by
\be \label{Laplacian}\Delta = 4\tau_2^2 \frac{\p^2}{\p\tau \p\bar\tau}.\ee
The Eisenstein series $E_k$ satisfies the Laplace equation
\be \label{LapE}\Delta E_k = k(k-1) E_k\ee 
while the modular graph $D_4$ satisfies the Poisson equation~\cite{DHoker:2015gmr,DHoker:2016mwo,Basu:2016xrt} 
\be \label{D4}\Big(\Delta -2\Big)\Big(D_4 - 3 E_2^2\Big) = 36 E_4 - 24 E_2^2.\ee
Also the graph $C_{1,1,2}$ satisfies the non--trivial relation~\cite{DHoker:2015gmr,DHoker:2015sve,DHoker:2016mwo,Basu:2016xrt}
\be \label{rel}\frac{1}{24}D_4 = C_{1,1,2} -\frac{3}{4} E_4 +\frac{1}{8}E_2^2\ee
hence relating it to the other graphs having distinct topologies. 

Next let us consider the elliptic modular graph functions. They are real--valued functions of $\tau$, the complex structure modulus of the torus, and $v \in \S$. Also they are invariant under the $SL(2,\mathbb{Z})$ transformation
\be \tau \rightarrow \frac{a\tau+b}{c\tau+d}, \quad v \rightarrow \frac{v}{c\tau+d},\ee
where $a,b,c,d \in \mathbb{Z}$ and $ad-bc=1$. The ones that are relevant for our purposes are given in figure 2. In this as well in later figures, the unintegrated vertices are labelled by $v$ and 0, while the remaining vertices are integrated\footnote{The labels $v$ and $0$ for the unintegrated vertices can be interchanged, which follows from elementary properties of the Green function.}.    

\begin{figure}[ht]
\begin{center}
\[
\mbox{\begin{picture}(260,210)(0,0)
\includegraphics[scale=.6]{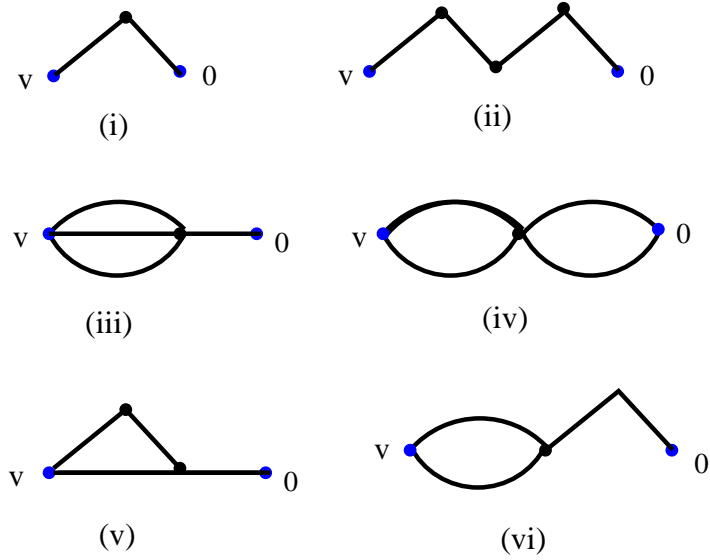}
\end{picture}}
\]
\caption{The elliptic modular graphs (i) $G_2 (v)$, (ii) $G_4 (v)$, (iii) $D_4^{(1)} (v)$, (iv) $D_4^{(2)} (v)$, (v) $D_4^{(1,2)} (v)$, (vi) $D_4^{(1,1,2)} (v)$}
\end{center}
\end{figure}

In figure 2, the first two graphs are special cases of the iterated Green function $G_k (v)$ ($k \geq 1$) which is defined recursively by 
\be G_{k+1} (v) = \int_{\S} \frac{d^2z}{\tau_2}G(v,z) G_k (z)\ee
where $G_1 (z) = G (z)$, the Green function. 
They satisfy the Laplace equation\footnote{Expressing $v$ as $v= x+ \tau y$, where $x, y \in (0,1]$, we see that $\frac{\p}{\p \tau}$ (and hence $\Delta$) acts trivially on $x,y$, though it acts non-trivially on functions of
$v$.}
\be \label{LapG}\Delta G_k (v) = k(k-1) G_k (v).\ee
The next two graphs in figure 2 are special cases of the elliptic graph ($k \leq l$)
\be D_l^{(k)} (v) = \int_{\S} \frac{d^2 z}{\tau_2}G^{k} (v,z)G (z)^{l-k} = D_l^{(l-k)} (v).\ee
Note that $D_l^{(k)} (0) = D_l$, where
\be D_l = \int_{\S}\frac{d^2 z}{\tau_2} G(z)^l,\ee 
of which $D_4$ in \C{defD} is a special case. Also note that $D_2 = E_2$.

Finally, the last two graphs in figure 2 is given by
\bea D_4^{(1,2)} (v) = \int_{\S^2} \frac{d^2 z}{\tau_2} \frac{d^2 w}{\tau_2} G(v,z)G(v,w)G(w,z)G(z), \quad D_4^{(1,1,2)} (v) = \int_{\S} \frac{d^2 z}{\tau_2} G(v,z)^2 G_2 (z) .\non \\ \eea

Note that
\be G_{k} (0) = E_k ~~(k \geq 2), \quad D_4^{(1,2)} (0) = D_4^{(1,1,2)} (0) = C_{1,1,2}.\ee

Our primary aim is to obtain Poisson equations satisfied by linear combinations of these graphs. To do so, we shall not directly apply \C{Laplacian} on the graphs as this is quite cumbersome. Instead, we shall analyze the variations of the graphs under complex structure deformations to obtain the Poisson equation, along the lines of~\cite{Basu:2015ayg,Basu:2016xrt,Basu:2016kli,Kleinschmidt:2017ege,Basu:2019idd}.     

The variations are given by~\cite{Verlinde:1986kw,DHoker:1988pdl,DHoker:2015gmr}
\be \label{Beltrami}\p_\mu G(z_1,z_2) = -\frac{1}{\pi}\int_{\S}d^2 z \p_z G(z,z_1) \p_z G(z,z_2),\ee
and
\be \label{Beltrami2}\overline\p_\mu\p_\mu G(z_1,z_2)=0\ee
where $\mu$ is the Beltrami differential.
Also the Laplacian is given in terms of these variations by
\be \Delta = \overline\p_\mu\p_\mu.\ee

Thus very briefly, the aim is to consider the $\overline\p_\mu \p_\mu$ variation of the graphs, integrate by parts appropriately using the single valued nature of the Green function, and simplify the resulting expressions using the equations
\bea \label{eigen}\overline\p_w\p_z G(z,w) &=& \pi \delta^2 (z-w) - \frac{\pi}{\tau_2}, \non \\
\overline\p_z\p_z G(z,w) &=& -\pi \delta^2 (z-w) + \frac{\pi}{\tau_2}\eea
satisfied by the Green function. Now the $\overline\p_\mu \p_\mu$ variation of any graph yields graphs having a maximum of two $\p$ and two $\overline\p$ operators acting on the Green function along its links using \C{Beltrami}, which we want to manipulate to obtain the Poisson equation.

\begin{figure}[ht]
\begin{center}
\[
\mbox{\begin{picture}(250,80)(0,0)
\includegraphics[scale=.7]{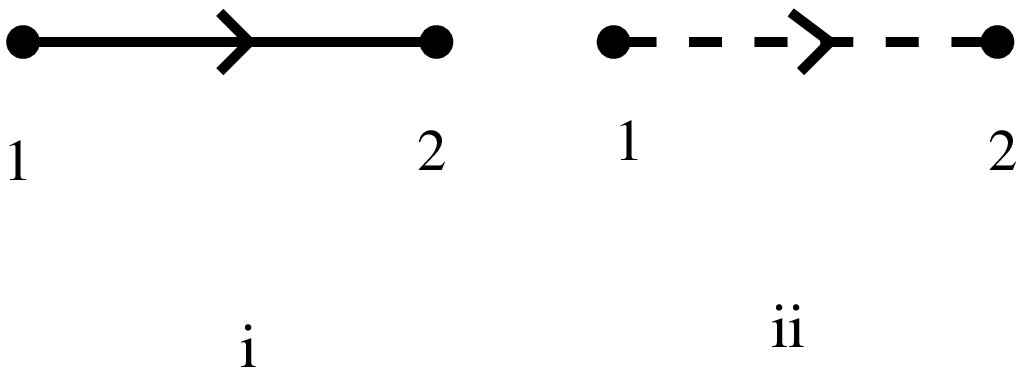}
\end{picture}}
\]
\caption{(i) $\p_{z_2} G(z_1,z_2) = -\p_{z_1} G(z_1,z_2)$,  (ii) $\overline\p_{z_2} G(z_1,z_2) = -\overline\p_{z_1} G(z_1,z_2)$}
\end{center}
\end{figure}

Derivatives of Green functions along the links will be denoted graphically as depicted by figures 3 and 4. 

\begin{figure}[ht]
\begin{center}
\[
\mbox{\begin{picture}(130,65)(0,0)
\includegraphics[scale=.6]{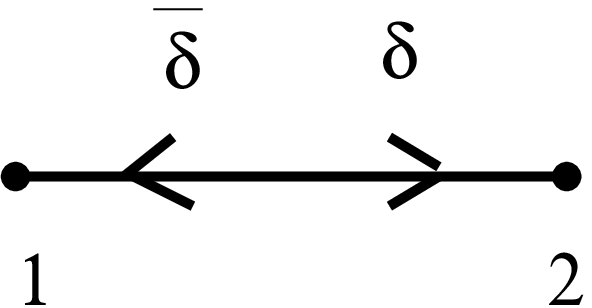}
\end{picture}}
\]
\caption{$\overline\p_{z_1} \p_{z_2} G(z_1,z_2)$}
\end{center}
\end{figure}

We now proceed to obtain the Poisson equations satisfied by the various graphs.

\section{Poisson equations for the elliptic modular graphs}

To motivate the choices of elliptic modular graphs that can potentially satisfy useful Poisson equations, consider the family of modular invariants~\cite{DHoker:2018mys} 
\be \label{defF}F_{2k} (v) = \frac{1}{(2k)!} \int_{\S} \frac{d^2 z}{\tau_2} f(z)^{2k},\ee
where $k$ is a positive integer, and $f(z) = G(v,z) -G(z)$.

The simplest case is given by $F_2 (v)$ which yields
\be \label{F2}F_2 (v) = E_2 - G_2 (v)\ee
by analyzing \C{defF} directly.
Thus $F_2 (v)$ satisfies the Laplace equation
\be \Big(\Delta -2\Big) F_2 (v) = 0\ee
which follows from \C{LapE} and \C{LapG}.

We now want to analyze the Poisson equation satisfied by $F_4 (v)$. From \C{defF}, we get that 
\be \label{F4}F_4 (v)= \frac{D_4}{12} -\frac{D_4^{(1)}(v)}{3} +\frac{D_4^{(2)}(v)}{4}\ee
in terms of the various graphs listed earlier. 

In the various expressions to follow, for the sake of brevity, we shall denote an integrated vertex over the location $z_i$ as $\int_{i}$. The measure can be either $d^2 z_i/\tau_2$ or $d^2 z_i$. While the factor of $\tau_2$ arises from the original graphs, the lack of this factor in the integration measure follows from the variation \C{Beltrami}. At the very end, we shall obtain all graphs with the measure factor $d^2z_i/\tau_2$ for every integrated vertex, which is a consequence of modular invariance.       

In order to consider the action of the Laplacian $\Delta$ on $F_4 (v)$ in \C{F4}, we note that its action on $D_4$ is given by \C{D4}, and thus we need to consider its action on the remaining two graphs in \C{F4}.

Using \C{Beltrami} and \C{Beltrami2}, we see that the action of $\Delta$ on $D_4^{(1)} (v)$ is given by
\bea \label{one}\Delta D_4^{(1)} (v) &=& 6 \int_1 G(v,z_1) \p_\mu G(v,z_1) \overline\p_\mu G(v,z_1) G(z_1) \non \\ &&+ 3\Big[\int_1 G(v,z_1)^2\p_\mu G(v,z_1) \overline\p_\mu G(z_1)+c.c.\Big],\eea
where the first term is manifestly real. Integrating by parts and using \C{eigen} repeatedly, we get that the first term in \C{one} yields\footnote{The various graphs $H_i (v), K_i (v)$ and $P_i (v)$ needed in our analysis are given in the appendix, and also depicted by figures 5, 6 and 7.}
\bea \label{1}\int_1 G(v,z_1) \p_\mu G(v,z_1) \overline\p_\mu G(v,z_1) G(z_1) &=& D_4^{(1,2)}(v) - \frac{G_2(v)^2}{2} + E_4 + C_{1,1,2} - \frac{E_2^2}{2} \non \\ &&-\frac{H_1(v)}{\pi} - \frac{1}{2\pi}\Big(H_2(v) + c.c.\Big) +\frac{1}{\pi^2}\Big(K_1 (v)+ c.c.\Big).\non \\ \eea

\begin{figure}[ht]
\begin{center}
\[
\mbox{\begin{picture}(340,280)(0,0)
\includegraphics[scale=.75]{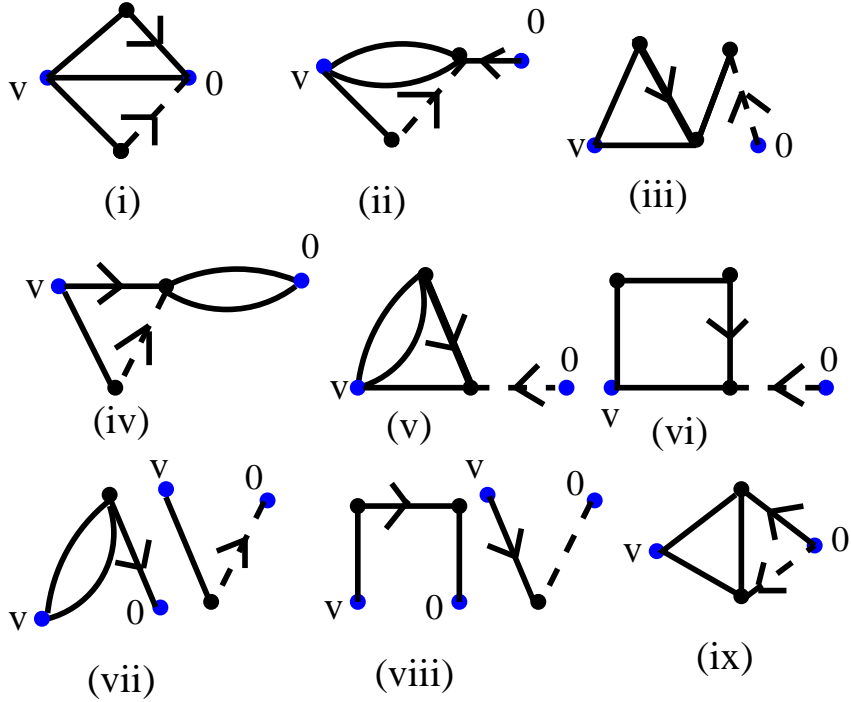}
\end{picture}}
\]
\caption{The graphs (i) $H_1 (v)$, (ii) $H_2 (v)$, (iii) $H_3 (v)$, (iv) $H_4 (v)$, (v) $H_5 (v)$, (vi) $H_6 (v)$, (vii) $H_7 (v)$, (viii) $H_8 (v)$, (ix) $H_9 (v)$}
\end{center}
\end{figure}

In obtaining equations like this, and also later, we use elementary relations like
\bea &&\int_{1,2,3} G(v,z_1)G(v,z_2)\p_{z_3} G(z_1,z_3)\overline\p_{z_3} G(z_2,z_3) G(z_3)= \pi D_4^{(1,2)}(v) - \frac{\pi}{2}G_2(v)^2 +\frac{\pi}{2}E_4, \non \\ &&\int_{1,2,3} G(z_1)G(z_2)G(z_3)\p_{z_3} G(z_1, z_3)\overline\p_{z_3}G(z_2, z_3) = \pi C_{1,1,2} - \frac{\pi}{2}E_2^2 +\frac{\pi}{2}E_4\eea
to express integrals involving derivatives of Green functions in term of modular graphs as well as their elliptic cousins.   

Proceeding similarly, the second term in \C{one} gives
\bea \label{2}\int_1 G(v,z_1)^2\p_\mu G(v,z_1) \overline\p_\mu G(z_1) = -\frac{2}{3} D_4^{(1)}(v)+\frac{H_2^*(v)}{\pi} +\frac{2}{\pi} H_3 (v) + \frac{2}{\pi^2}K_2(v).\eea

Thus we see that the action of $\Delta$ on $D_4^{(1)} (v)$ not only produces graphs with links given by Green functions, but it also produces graphs with links involving derivatives of Green functions.   

\begin{figure}[ht]
\begin{center}
\[
\mbox{\begin{picture}(280,230)(0,0)
\includegraphics[scale=.6]{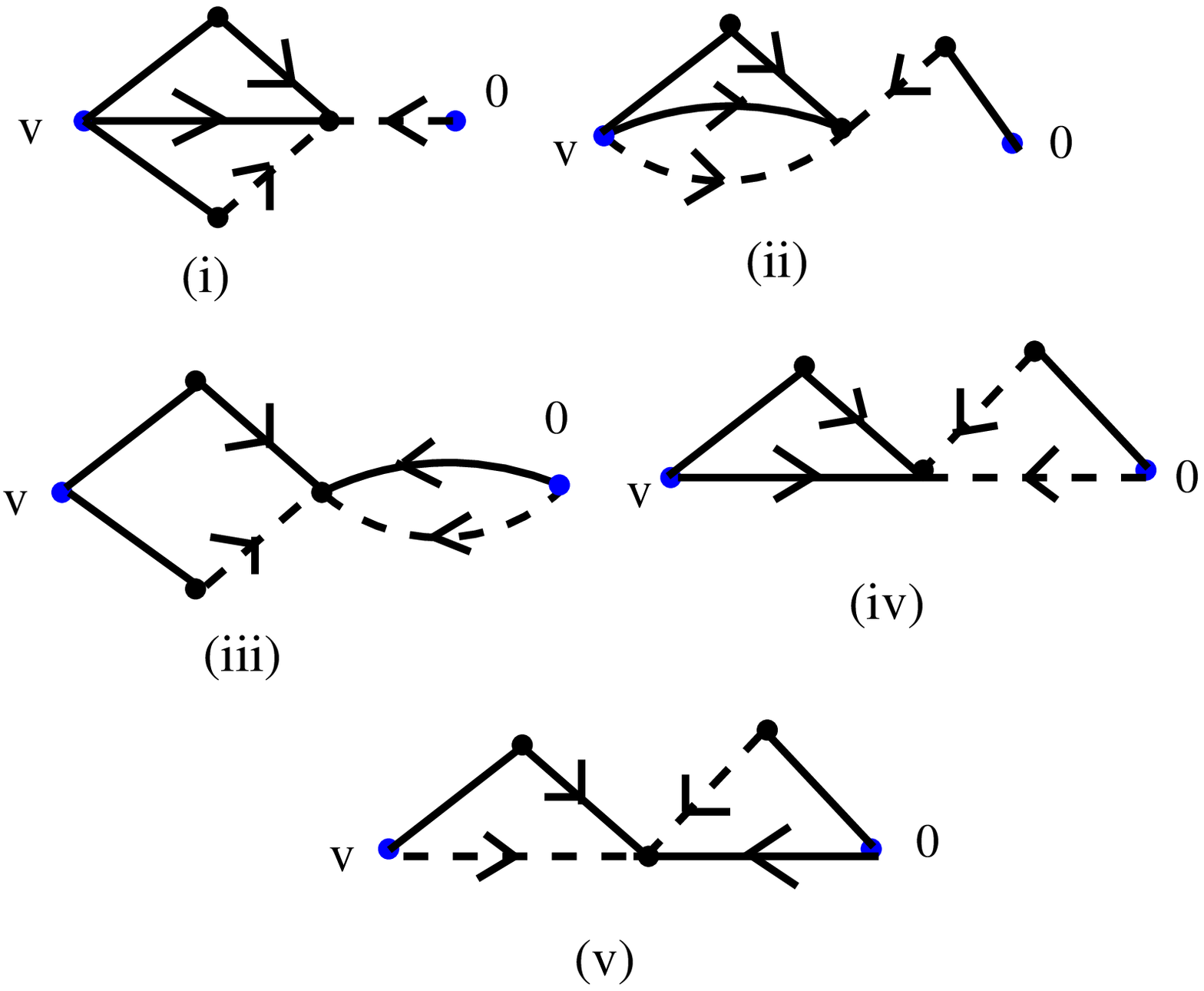}
\end{picture}}
\]
\caption{The graphs (i) $K_1 (v)$, (ii) $K_2 (v)$, (iii) $K_3 (v)$, (iv) $K_4 (v)$, (v) $K_5 (v)$}
\end{center}
\end{figure}

We now consider the action of $\Delta$ on $D_4^{(2)} (v)$ in \C{F4}. This gives us that
\bea \label{two}\Delta D_4^{(2)} (v) = 4 \int_1 \p_\mu G(v,z_1) \overline\p_\mu G(v,z_1) G(z_1)^2 + 8 \int_1 G(v,z_1) \p_\mu G(v,z_1) G(z_1) \overline\p_\mu G(z_1).\eea
Note that each contribution is manifestly real. 

Proceeding as before, the first and second terms in \C{two} yield
\bea \label{3}\int_1 \p_\mu G(v,z_1) \overline\p_\mu G(v,z_1) G(z_1)^2&=&- D_4^{(2)}(v) + 2 D_4^{(1,2)}(v) - G_2(v)^2 + E_4 \non \\ &&+\frac{1}{\pi}\Big(H_4 (v) + c.c.\Big) +\frac{2}{\pi^2} K_3 (v),  \eea
and
\bea \label{4} \int_1 G(v,z_1) \p_\mu G(v,z_1) G(z_1) \overline\p_\mu G(z_1) &=& \frac{1}{\pi}\Big(H_3 (v)+ c.c.\Big) -\frac{1}{2\pi}\Big(H_4 (v) + c.c.\Big) \non \\ &&+\frac{1}{\pi^2}\Big(K_4 (v) + K_5 (v)\Big),\eea
respectively.

Thus as with the case involving $D_4^{(1)}(v)$, we see that the action of $\Delta$ on $D_4^{(2)} (v)$ produces graphs with links given by Green functions, as well as graphs with links given by derivatives of Green functions.   

\begin{figure}[ht]
\begin{center}
\[
\mbox{\begin{picture}(280,300)(0,0)
\includegraphics[scale=.65]{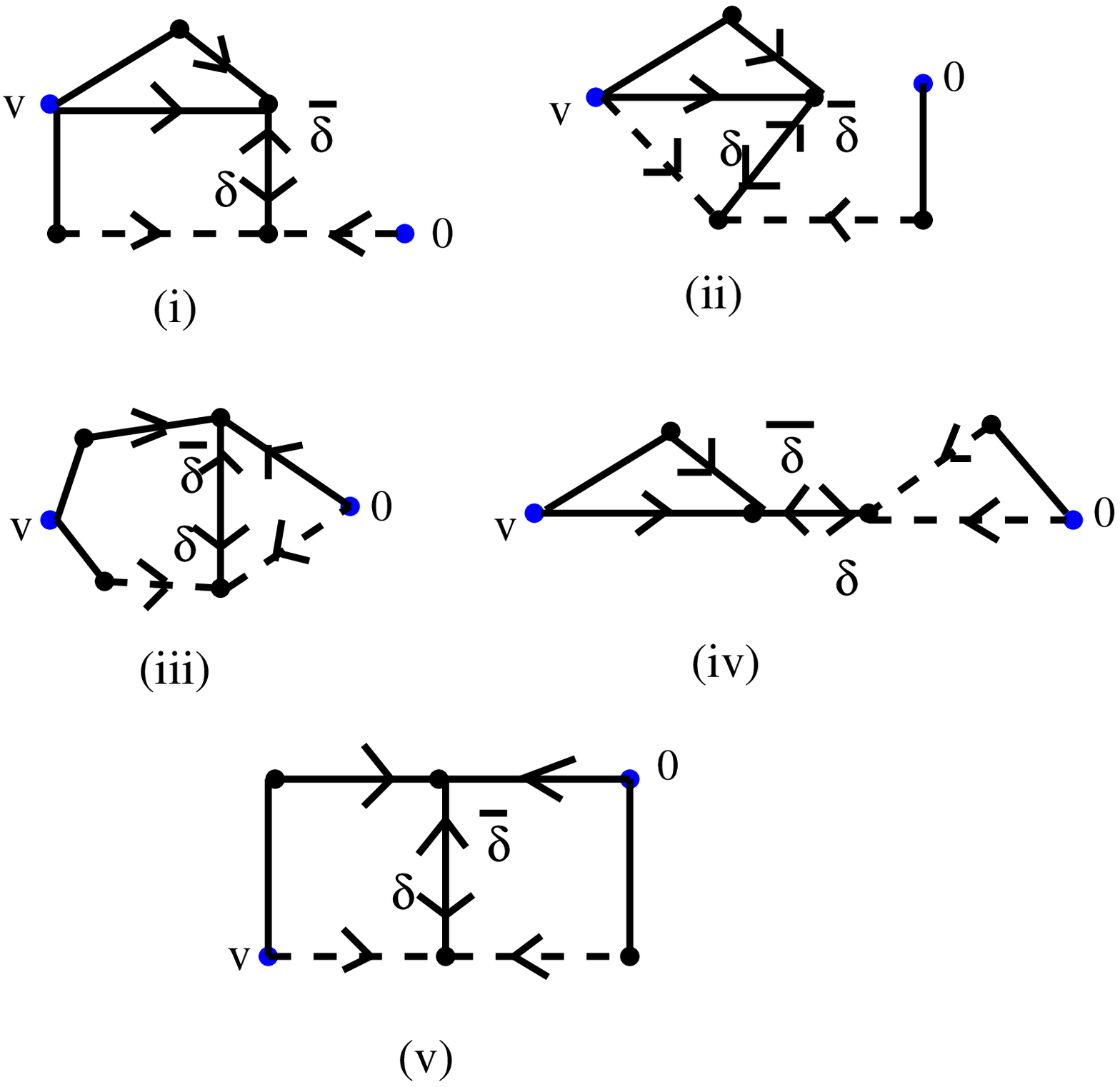}
\end{picture}}
\]
\caption{The auxiliary graphs (i) $P_1 (v)$, (ii) $P_2 (v)$, (iii) $P_3 (v)$, (iv) $P_4 (v)$, (v) $P_5 (v)$}
\end{center}
\end{figure}

Let us now consider the graphs $K_i (v)$ each of which has four derivatives of the Green function. 

To evaluate each of them, we start with an appropriate auxiliary graph~\cite{Basu:2016xrt}, which yields $K_i (v)$ on using \C{eigen} trivially. On the other hand the auxiliary graph is such that it can be independently evaluated on integrating by parts and using \C{eigen} such that it contains no contributions having more than two derivatives of the Green function. For every $K_i (v)$, the auxiliary graph is given by $P_i (v)$.    

Thus to evaluate $K_1 (v)$, we start with the auxiliary graph $P_1 (v)$. It can be simplified trivially by using \C{eigen} for the link which has both $\p$ and $\overline\p$ acting on the Green function, hence relating it to $K_1 (v)$. Alternatively, it can be evaluated by judiciously integrating by parts both $\p$ and $\overline\p$ which are on the same link, and hence moving the arrows around the circuit and using \C{eigen}. We see that this procedure only produces graphs having no more than two derivatives.        

For $K_1 (v)$, adding the contribution from the complex conjugate as well, we get that
\bea \label{K1}&&\Big( \frac{K_1 (v)}{\pi^2} -\frac{1}{4} \p_\mu E_2 \overline\p_\mu G_2 (v)\Big) + c.c. = 2 E_4 -  C_{1,1,2} + \frac{1}{2\pi}\Big(H_5 (v) + c.c.\Big) \non \\ &&-\frac{1}{\pi}\Big(H_6 (v) + c.c.\Big) +\frac{1}{2\pi}\Big(H_7 (v) + c.c.\Big) - \frac{1}{\pi}\Big(H_8 (v) + c.c.\Big).\eea

Proceeding similarly, for $K_2 (v)$, we get that 
\bea \label{K2}&&\Big( \frac{K_2 (v)}{\pi^2} -\frac{1}{4} \p_\mu E_2 \overline\p_\mu G_2 (v)\Big) + c.c. = D_4^{(1)} (v) - E_2 G_2 (v) + 2 G_4 (v) - D_4^{(1,1,2)} (v) \non \\ &&-2 D_4^{(1,2)} (v) -\frac{1}{2\pi}\Big(H_5 (v) + c.c.\Big) +\frac{1}{\pi}\Big(H_6 (v) + c.c.\Big).\eea

For $K_3 (v)$, $K_4 (v)$ and $K_5 (v)$, a similar analysis yields 
\bea \label{Krest} &&\frac{K_3 (v)}{\pi^2} -\frac{1}{4} \p_\mu G_2 (v) \overline\p_\mu G_2 (v) = E_4 +\frac{H_9 (v)}{\pi} - \frac{1}{\pi} \Big(H_6 (v) + c.c.\Big) \non \\ &&- \frac{1}{2\pi}\Big(H_7 (v) + c.c.\Big)-\frac{1}{\pi}\Big(H_8 (v) + c.c.\Big) , \non \\ 
&&\frac{K_4 (v)}{\pi^2} -\frac{1}{4} \p_\mu E_2 \overline\p_\mu E_2  = \frac{D_4^{(2)}(v)}{4} - \frac{E_2^2}{4} - D_4^{(1,1,2)} (v) + G_4 (v), \non \\ 
&& \frac{K_5 (v)}{\pi^2} -\frac{1}{4} \p_\mu G_2 (v) \overline\p_\mu G_2 (v) = -2 D_4^{(1,2)} (v) +\frac{1}{2}(D_4^{(2)}(v) - G_2 (v)^2) + G_4 (v) \non \\ &&-\frac{1}{\pi} \Big(H_1 (v) + H_9(v)\Big) +\frac{1}{\pi}\Big(H_6 (v) + c.c.\Big) +\frac{1}{\pi}\Big(H_7 (v) + c.c.\Big).\eea

Now let us consider the contributions to $\Delta D_4^{(1)} (v)$ and $\Delta D_4^{(2)}(v)$ that arise only from $H_i (v)$. These are given by terms involving $H_i (v)$ explicitly in \C{1}, \C{2}, \C{3} and \C{4}, as well as from  $K_i (v)$ that we obtain from the right hand side of \C{K1}, \C{K2} and \C{Krest}. Defining
\be \Lambda (v) = -\frac{6}{\pi} H_1 (v) +\frac{6}{\pi}\Big(H_3 (v) + c.c.\Big) +\frac{3}{\pi}\Big(H_7 (v) + c.c.\Big) - \frac{6}{\pi}\Big(H_8 (v) + c.c.\Big),\ee
we have that
\bea \label{L}\Delta D_4^{(1)} (v) - 3\Big(\p_\mu E_2 \overline\p_\mu G_2 (v) + c.c.\Big) &=& \Lambda (v) +\ldots, \non \\ \Delta D_4^{(2)} (v) - 4\p_\mu G_2 (v) \overline\p_\mu G_2 (v) - 2\p_\mu E_2 \overline\p_\mu E_2 &=&\frac{4}{3}\Lambda (v) +\ldots.\eea
where the remaining contributions on the right hand side do not contain derivatives and all other contributions involving $H_i (v)$ cancel. Strikingly, the right hand side of both the equations in \C{L} is proportional to $\Lambda$. Thus eliminating $\Lambda$ between the two equations, we get that
\bea \label{elim}&&\Delta \Big(D_4^{(1)}(v) - \frac{3}{4} D_4^{(2)}(v)\Big) -3\Big(\p_\mu E_2 \overline\p_\mu G_2 (v) + c.c.\Big) +3 \p_\mu G_2 (v) \overline\p_\mu G_2 (v) +\frac{3}{2} \p_\mu E_2 \overline\p_\mu E_2 \non \\ &&= 2\Big(D_4^{(1)} (v) - \frac{3}{4} D_4^{(2)}(v)\Big) + 9 E_4 + 3 G_2 (v)^2 - \frac{3}{2} E_2^2 -6 E_2 G_2 (v).\eea

Now the terms other than $\Delta D_4^{(1)} (v)$ and $\Delta D_4^{(2)} (v)$ on the left hand side of \C{elim} can be expressed in terms of the action of $\Delta$ on bilinears of graphs, which follows from the identities 
\bea \Delta \Big(E_2 G_2 (v)\Big) &=& 4 E_2 G_2 (v) +\Big(\p_\mu E_2 \overline\p_\mu G_2 (v) + c.c.\Big), \non \\ \Delta G_2(v)^2 &=& 4 G_2 (v)^2 + 2 \p_\mu G_2 (v) \overline\p_\mu G_2 (v), \non \\ \Delta E_2^2 &=& 4 E_2^2 + 2  \p_\mu E_2 \overline\p_\mu E_2 .\eea
This leads to the Poisson equation
\bea \label{P}&&\Delta \Big(D_4^{(1)}(v) - \frac{3}{4} D_4^{(2)}(v) - 3 E_2 G_2 (v) +\frac{3}{2} G_2(v)^2 +\frac{3}{4} E_2^2\Big) \non \\ &&= 2\Big(D_4^{(1)} (v) - \frac{3}{4} D_4^{(2)}(v)\Big) + 9 E_4 - 18 E_2 G_2 (v) + 9 G_2 (v)^2 +\frac{3}{2} E_2^2\eea
involving the various modular graphs and their elliptic generalizations.

Coupled with the Poisson equation for $D_4$ in \C{D4}, and the definitions \C{F2} and \C{F4} for $F_2 (v)$ and $F_4 (v)$ respectively, we see that \C{P} leads to the Poisson equation
\be \Delta\Big(F_4 (v) - \frac{1}{2} F_2(v)^2\Big) = 2 F_4(v) - 3 F_2(v)^2\ee
which is precisely the equation obtained in~\cite{DHoker:2020tcq}, hence providing a derivation directly at genus one. 

Note that the elliptic modular graphs $D_4^{(1,2)}(v)$ and $D_4^{(1,1,2)} (v)$ given in figure 2, that arose in the intermediate steps of the calculation, cancel in the final answer. While $D_4^{(1,2)} (v)$ does not yield a Poisson equation involving only graphs having links given by Green function and not their derivatives, $D_4^{(1,1,2)} (v)$ does yield such an equation.   

Proceeding along the lines of our previous analysis, we get that the Poisson equation satisfied by $D_4^{(1,1,2)} (v)$ is given by 
\be \Big(\Delta-2\Big) D_4^{(1,1,2)} (v) = 10 G_4 (v) - E_2^2 - E_4 +F_2(v)^2.\ee

\section{An algebraic relation among modular graphs}

Let us rewrite the Poisson equations satisfied by $D_4^{(1,1,2)}$, $F_4 (v)$ and $D_4$ keeping only $F_2(v)^2$ and $E_2^2$ as the source terms and absorbing all other contributions in the definition of the eigenfunction. This gives us that 
\bea &&\Big(\Delta -2\Big) \Big( D_4^{(1,1,2)} (v) - G_4 (v) +\frac{1}{10}E_4\Big) = F_2 (v)^2 - E_2^2,\non \\
&&\Big(\Delta -2\Big)\Big(F_4 (v) - \frac{1}{2} F_2 (v)^2\Big) = - 2 F_2 (v)^2, \non \\ &&\Big(\Delta -2\Big) \Big(D_4 - 3 E_2^2 - \frac{18}{5} E_4\Big) = -24 E_2^2.\eea
Note that each equation has the same eigenvalue.

\begin{figure}[ht]
\begin{center}
\[
\mbox{\begin{picture}(420,140)(0,0)
\includegraphics[scale=.67]{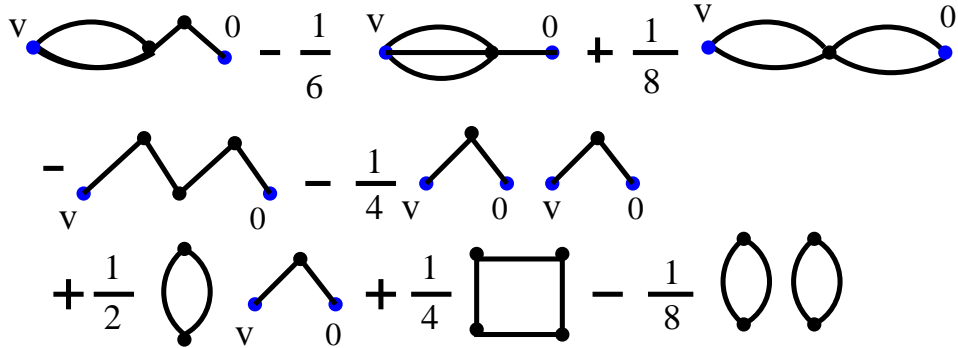}
\end{picture}}
\]
\caption{The graphical expression for $\Phi (v)$}
\end{center}
\end{figure}

Eliminating the source terms between these equations, we get that 
\be \label{E}\Big(\Delta-2\Big) \Phi(v) = 0,\ee
where
\bea \Phi(v) = D_4^{(1,1,2)} (v) - \frac{1}{6}D_4^{(1)}(v) +\frac{1}{8} D_4^{(2)}(v) - G_4 (v) -\frac{1}{4}G_2(v)^2 +\frac{1}{2}E_2 G_2 (v) +\frac{1}{4}E_4 - \frac{1}{8} E_2^2, \non \\ \eea
which is denoted graphically in figure 8.
To solve \C{E} consistent with $SL(2,\mathbb{Z})$ invariance and power law growth as the boundary condition for large $\tau_2$, we take the ansatz
\be \label{Phi}\Phi(v) = \alpha E_2 +\beta G_2 (v),\ee
where $\alpha$ and $\beta$ are constants\footnote{It would be interesting to prove this statement, perhaps along the lines of~\cite{Terras1}.}. We expect $\alpha$ and $\beta$ to vanish simply because every term on the left hand side of \C{Phi} has four links, while those on the right hand side have two links.   

To see this, we first integrate over the vertex marked 0 over $\S$ in figure 8, with the measure $d^2 z/\tau_2$. Using \C{Vanish}, we see the integrals of $D_4^{(1,1,2)}(v), D_k^{(1)} (v)$ and $G_k (v)$ vanish, while the integrals of $D_4^{(2)}(v)$ and $G_2(v)^2$ yield $E_2^2$ and $E_4$ respectively. Thus the left hand side of \C{Phi} trivially vanishes as contributions that arise involving $E_4$ and $E_2^2$ cancel, while the right hand side yields $\alpha E_2$, thus yielding $\alpha =0$. 

Next we identify the vertices marked $v$ and 0 in figure 8, resulting in modular graphs where all vertices are integrated. Setting $\alpha =0$, from \C{Phi} we get that
\be \label{van}C_{1,1,2} -\frac{1}{24} D_4 -\frac{3}{4}E_4 +\frac{1}{8} E_2^2 = \beta E_2.\ee
Using the identity \C{rel} the left hand side of \C{van} vanishes, and hence $\beta =0$. Thus 
\be \Phi(v) =0, \ee
which yields a non--trivial algebraic relation between several elliptic modular graphs and modular graphs of distinct topologies.  

Note that among elliptic graphs with four links, the two graphs in figure 9 are trivially equal, which easily follows along the lines of our analysis.   
\begin{figure}[ht]
\begin{center}
\[
\mbox{\begin{picture}(285,40)(0,0)
\includegraphics[scale=.67]{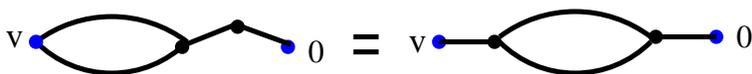}
\end{picture}}
\]
\caption{An equality among elliptic graphs}
\end{center}
\end{figure}

Thus we see that the analysis of the variations of the graphs under complex structure deformations can be manipulated to yield Poisson equations they satisfy. The structure of the eigenfunctions follows naturally, where the various subtractions from the parent set of graphs arise from manipulations involving the auxiliary graphs. It will be interesting to generalize this procedure to have a detailed understanding of Poisson equations satisfied by various elliptic modular graphs. It will also be interesting to understand the possible origin of these Poisson equations from identities among modular graphs at genus two, where they should arise in the asymptotic expansion around the non--separating node.

\appendix

\section{The list of various graphs}

In this appendix, we list the various graphs that are needed in the main text. They are depicted graphically in figures 5, 6 and 7.

Each of the graphs $H_i (v)$ $(i=1,\ldots,9)$ has two derivatives of the Green function. They are given by
\bea H_1 (v) &=& \int_{1,2} G(v,z_1) G(v,z_2) G(v) \p_{z_1} G(z_1) \overline\p_{z_2} G(z_2), \non \\ H_2 (v) &= &\int_{1,2} G(v,z_1)^2 G(v,z_2) \p_{z_1} G(z_1) \overline\p_{z_1} G(z_1,z_2), \non \\ H_3 (v) &= &\int_{1,2,3} G(v,z_1) G(v,z_2) G(z_2,z_3) \p_{z_2} G(z_1,z_2) \overline\p_{z_3} G(z_3), \non \\ H_4 (v) &=& \int_{1,2} G(v,z_1) G(z_2)^2 \p_{z_2} G(v,z_2) \overline\p_{z_2} G(z_1,z_2),\non \\ H_5 (v) &=& \int_{12} G(v,z_1)^2 G(v,z_2) \p_{z_2} G(z_1,z_2) \overline\p_{z_2} G(z_2), \non \\ H_6 (v) &=& \int_{123} G(v,z_1)G(z_1,z_2) G(v,z_3) \p_{z_3} G(z_2,z_3) \overline\p_{z_3} G(z_3), \non \\ H_7 (v)  &=& \int_1 G(v,z_1)^2 \p_{z_1} G(z_1) \int_{z_2} G(v,z_2) \overline\p_{z_2} G(z_2),\non \\ H_8 (v) &=& \int_{1,2} G(v,z_1)G(z_2) \p_{z_2}G(z_1,z_2) \int_3 G(z_3)\overline\p_{z_3} G(v,z_3), \non \\ H_9(v) &=& \int_{1,2} G(v,z_1) G(v,z_2) G(z_1,z_2) \p_{z_1} G(z_1) \overline\p_{z_2} G(z_2).
\eea
 
The graphs $K_i (v)$ $(i=1,\ldots,5)$, each of which has four derivatives of the Green function, are given by
\bea K_1 (v) &=& \int_{1,2,3} G(v,z_1) G(v,z_2) \p_{z_3}G(z_1,z_3) \p_{z_3} G(v,z_3) \overline\p_{z_3} G(z_2,z_3) \overline\p_{z_3} G(z_3), \non \\ K_2 (v) &=& \int_{1,2,3} G(v,z_1) G(z_3)\p_{z_2} G(z_1,z_2) \p_{z_2}G(v,z_2) \overline\p_{z_2} G(v,z_2) \overline\p_{z_2}G(z_2,z_3) , \non \\ K_3 (v) &=& \int_{1,2,3} G(v,z_1)G(v,z_2) \p_{z_3}G(z_1,z_3) \p_{z_3}G(z_3) \overline\p_{z_3}G(z_2,z_3) \overline\p_{z_3}G(z_3),\non \\ K_4 (v) &=& \int_{1,2,3} G(z_3) G(v,z_1) \p_{z_2} G(z_1,z_2) \p_{z_2}G(v,z_2) \overline\p_{z_2} G(z_2,z_3) \overline\p_{z_2}G(z_2), \non \\ K_5 (v) &=& \int_{1,2,3} G(z_3) G(v,z_1) \p_{z_2}G(z_1,z_2) \p_{z_2}G(z_2) \overline\p_{z_2} (v,z_2) \overline\p_{z_2} G(z_2,z_3).\eea

Finally, the auxiliary graphs $P_i (v)$ ($i=1,\ldots,5$) needed to simplify the graphs involving $K_i (v)$ are given by 

\bea P_1 (v) &=& \int_{1,2,3,4} G(v,z_1) G(v,z_4)\p_{z_2} G(z_1,z_2) \p_{z_2} G(v,z_2) \overline\p_{z_3}G(z_3) \overline\p_{z_3} G(z_3,z_4) \overline\p_{z_2} \p_{z_3} G(z_2,z_3), \non \\ P_2 (v) &=& \int_{1,2,3,4} G(v,z_1) G(z_4) \p_{z_2} G(z_1,z_2) \p_{z_2} G(v,z_2)  \overline\p_{z_3} G(v,z_3) \overline\p_{z_3} G(z_3,z_4) \overline\p_{z_2} \p_{z_3} G(z_2,z_3), \non \\ P_3 (v) &=& \int_{1,2,3,4} G(v,z_1) G(v,z_2)\p_{z_3} G(z_1,z_3) \p_{z_3} G(z_3) \overline\p_{z_4} G(z_2,z_4) \overline\p_{z_4} G(z_4) \overline\p_{z_3} \p_{z_4} G(z_3,z_4), \non \\ P_4 (v) &=& \int_{1,2,3,4} G(v,z_1) G(z_4) \p_{z_2} G(z_1,z_2) \p_{z_2} G(v,z_2) \overline\p_{z_3} G(z_3,z_4) \overline\p_{z_3} G(z_3) \overline\p_{z_2} \p_{z_3} G(z_2,z_3), \non \\ P_5 (v) &=& \int_{1,2,3,4} G(v,z_1) G(z_3) \p_{z_2} G(z_1,z_2) \p_{z_2} G(z_2) \overline\p_{z_3}G(z_3,z_4) \overline\p_{z_4} G(v,z_4) \overline\p_{z_2}\p_{z_4} G(z_2,z_4).\non \\ \eea

\providecommand{\href}[2]{#2}\begingroup\raggedright\endgroup

%\bibliographystyle{utphys}
%\bibliography{myrefs}

\end{document}